\documentclass[12pt]{article}
\usepackage{graphicx}
\usepackage{epsfig}
\usepackage{amsfonts}

\usepackage{a4}

\usepackage{amsmath}

\usepackage{epsfig}
 
\newcommand{\beq}{\begin{equation}}
\newcommand{\eeq}{\end{equation}}
\newcommand{\beqs}{\begin{eqnarray}}
\newcommand{\eeqs}{\end{eqnarray}}
\begin{document}
\title{Magnetic solutions in AdS$_5$ and trace anomalies}
\author{{\large Yves Brihaye}$^{\dagger}$ and
{\large Eugen Radu}$^{\ddagger}$
 \\ 
$^{\dagger}${\small Physique-Math\'ematique, Universite de
Mons-Hainaut, Mons, Belgium}\\  
$^{\ddagger}${\small Department of
Mathematical Physics, National University of Ireland Maynooth,} \\
{\small Maynooth, Ireland}  }

\date{}
\newcommand{\dd}{\mbox{d}}
\newcommand{\tr}{\mbox{tr}}
\newcommand{\la}{\lambda}
\newcommand{\ka}{\kappa}
\newcommand{\f}{\phi}
\newcommand{\al}{\alpha}
\newcommand{\ga}{\gamma}
\newcommand{\de}{\delta}
\newcommand{\si}{\sigma}
\newcommand{\bomega}{\mbox{\boldmath $\omega$}}
\newcommand{\bsi}{\mbox{\boldmath $\sigma$}}
\newcommand{\bchi}{\mbox{\boldmath $\chi$}}
\newcommand{\bal}{\mbox{\boldmath $\alpha$}}
\newcommand{\bpsi}{\mbox{\boldmath $\psi$}}
\newcommand{\brho}{\mbox{\boldmath $\varrho$}}
\newcommand{\beps}{\mbox{\boldmath $\varepsilon$}}
\newcommand{\bxi}{\mbox{\boldmath $\xi$}}
\newcommand{\bbeta}{\mbox{\boldmath $\beta$}}
\newcommand{\ee}{\end{equation}}
\newcommand{\eea}{\end{eqnarray}}
\newcommand{\be}{\begin{equation}}
\newcommand{\bea}{\begin{eqnarray}}
\newcommand{\ii}{\mbox{i}}
\newcommand{\e}{\mbox{e}}
\newcommand{\pa}{\partial}
\newcommand{\Om}{\Omega}
\newcommand{\vep}{\varepsilon}
\newcommand{\bfph}{{\bf \phi}}
\newcommand{\lm}{\lambda}
\def\theequation{\arabic{equation}}
\renewcommand{\thefootnote}{\fnsymbol{footnote}}
\newcommand{\re}[1]{(\ref{#1})}
\newcommand{\R}{{\rm I \hspace{-0.52ex} R}}
\newcommand{\N}{{\sf N\hspace*{-1.0ex}\rule{0.15ex}%
{1.3ex}\hspace*{1.0ex}}}
\newcommand{\Q}{{\sf Q\hspace*{-1.1ex}\rule{0.15ex}%
{1.5ex}\hspace*{1.1ex}}}
\newcommand{\C}{{\sf C\hspace*{-0.9ex}\rule{0.15ex}%
{1.3ex}\hspace*{0.9ex}}}
\newcommand{\eins}{1\hspace{-0.56ex}{\rm I}}
\renewcommand{\thefootnote}{\arabic{footnote}}

\maketitle


\begin{abstract}
We discuss black hole and black string solutions 
in $d=5$ Einstein-Yang-Mills theory with negative cosmological
constant, 
proposing a  method to compute their  mass and action.
The magnetic gauge field of these configurations 
does not vanish at infinity. 
We argue that this implies a nonvanishing trace for the
stress tensor of the dual $d=4$ theory.
\end{abstract}
\medskip 
 
\noindent{\textbf{Introduction.--~}}As originally found in $d=4$ spacetime dimensions
\cite{Winstanley:1998sn}, \cite{Bjoraker:2000qd},
a variety of well known features of asymptotically flat self-gravitating non-Abelian
solutions are not shared by their anti-de Sitter (AdS) counterparts.
In the presence of a negative cosmological constant $\Lambda<0$, the
Einstein-Yang-Mills (EYM) theory 
possesses a continuum spectrum  of  regular and black hole non-Abelian   
solutions in terms of the adjustable  parameters
that specifies the initial conditions at the origin or at the event horizon,  
rather then discrete points.
The gauge field of generic solutions does not vanish asymptotically,
resulting in a nonzero magnetic flux at infinity.
Moreover, in contrast with the $\Lambda=0$ case,
some of the AdS configurations are stable against  linear perturbations 
\cite{Breitenlohner:2003qj}.
As found in \cite{Okuyama:2002mh}, \cite{Radu:2005mj} these features are shared 
by higher dimensional spherically symmetric AdS non-Abelian solutions.

Since gauged supergravity theories  generically contain non-Abelian matter fields in
the bulk,  these configurations are relevant in an AdS/CFT context, 
 offering the possibility
of studying some aspects of the nonperturbative structure
of a CFT in a background gauge field  \cite{Mann:2006jc}.
 On the CFT side, the boundary non-Abelian fields correspond  to external source currents
 coupled to various operators.

However, in contrast with the 
four dimensional case, 
a generic property of $d>4$  non-Abelian solutions is that their mass and action,
as defined in the usual way, diverge  \cite{Okuyama:2002mh,Radu:2005mj},
which may raise questions
about their physical relevance.
For example, in the best understood $d=5$ case \cite{Okuyama:2002mh}, 
 although the spacetime still
approaches asymptotically the maximally symmetric background,
the total action presents a logarithmically
divergent part.
The  coefficient of the divergent term is proportional to the square of the induced 
non-Abelian field on the boundary at infinity\footnote{The 
existence of a logarithmic divergence in the action
is a known property of some classes 
of AdS$_5$ solutions with a special boundary geometry  \cite{Emparan:1999pm}. 
The coefficients of the divergent terms there are related to the conformal 
Weyl anomaly in the dual theory \cite{Skenderis:2000in,odintsov}.
However, this is not the case of the non-Abelian AdS$_5$ configurations in 
\cite{Okuyama:2002mh}, which have the same boundary metric
as the Schwarzschild-AdS (SAdS) solution and thus no Weyl anomaly in the dual CFT.}.

Here we argue that the logarithmic divergence of the  
non-Abelian AdS$_5$ configurations does not signal
a problem with these solutions,
but rather provides a consistency check of the AdS/CFT conjecture.
The coefficient of the divergent term in the action
is related in this case to the trace anomaly of the dual CFT
defined in a background non-Abelian magnetic field. 
In this context, we propose to compute the mass and action of these 
solutions by using a counterterm prescription.
This enables us to discuss the thermodynamical properties
of two classes of AdS$_5$ non-Abelian black objects.
 
\noindent{\textbf{Non-Abelian black hole solutions.--~}}The
action of the $d=5$
gauged
supergravities usually contain the YM term 
$L_{YM}=-1/(2e^2){\rm Tr}\{F_{\mu \nu }F^{\mu \nu} \}$ 
as a basic building block (with $F_{\mu \nu}$ the field strength 
 and $e$
the gauge coupling constant).
In what follows we
consider a truncation of such models corresponding
to a pure EYM theory  with a lagrangean 
density\footnote{Usually, 
one has also to consider 
a non-Abelian Chern-Simon term. However, for purely magnetic solutions discussed here,
this term vanishes identically.} 
$L=1/(16\pi G)(R-2\Lambda)+L_{YM}$, with $\Lambda=-6/\ell^2$. 

The first class of solutions we consider corresponds to 
spherically symmetric or topological black holes with a metric ansatz
\begin{eqnarray}
\label{metric-gen} 
ds^{2}= 
 \frac{dr^2}{N(r)}+ r^2d\Omega^2_{3,k}-  N(r)\sigma^2(r) dt^{2},
\end{eqnarray}
where $d\Omega^2_{3,k}=d\psi^{2}+f^{2}_k(\psi) 
(d \theta^2+\sin^2\theta d \varphi^2)$ denotes the line
element of a three-dimensional space $\Sigma$ with
constant curvature.
The discrete parameter $k$ takes the values $1, 0$ and $-1$
and implies the form of the function $f_k(\psi)$: 
when $k=1$, $f_1(\psi)=\sin\psi$ and the hypersurface $\Sigma$ 
represents a 3-sphere; 
for $k=-1$,
it is a $3-$dimensional negative constant curvature space and
$f_{-1}(\psi)=\sinh\psi$.
The case $k=0$ is with $f_0(\psi)= \psi$ and $\Sigma$ a flat surface.

Restricting to an SU(2)
gauge field, the YM ansatz compatible with the symmetries 
of the line-element (\ref{metric-gen})
reads \cite{Okuyama:2004in}, \cite{Radu:2006va} (with $\tau_a$ the Pauli spin
matrices)
\begin{eqnarray} 
\label{A}
A=\frac{1}{2} \Big\{ 
\tau_3(\omega(r) d \psi +\cos \theta d \varphi)
-\frac{d f_k(\psi)}{d \psi}(\tau_2 d \theta+\tau_1 \sin
\theta d \varphi) +\omega(r)f_k(\psi)(\tau_1 d \theta-\tau_2 \sin
\theta d \varphi)
 \Big\}.
\end{eqnarray} 
The resulting set of three
ordinary differential equations is solved with suitable boundary conditions.
Supposing the existence of an event horizon for some $r_h>0$, one imposes
$N(r_h)=0$, $\sigma(r_h)=\sigma_h>0$, $w(r_h)=w_h$.
By going to the Euclidean section (or by computing the surface gravity)
one finds the black holes Hawking temperature
$T_H={1}/{\beta}={\sigma_h N'(r_h)}/{4 \pi}.$
(One should note that these non-Abelian magnetic solutions
extremize also the
Euclidean action, the Wick rotation 
$t \to it$ having no effect at the level of the equations of motion.)
\begin{figure}[h!]
\parbox{\textwidth}
{\centerline{
\mbox{
\epsfysize=10.0cm
\includegraphics[width=87mm,angle=0,keepaspectratio]{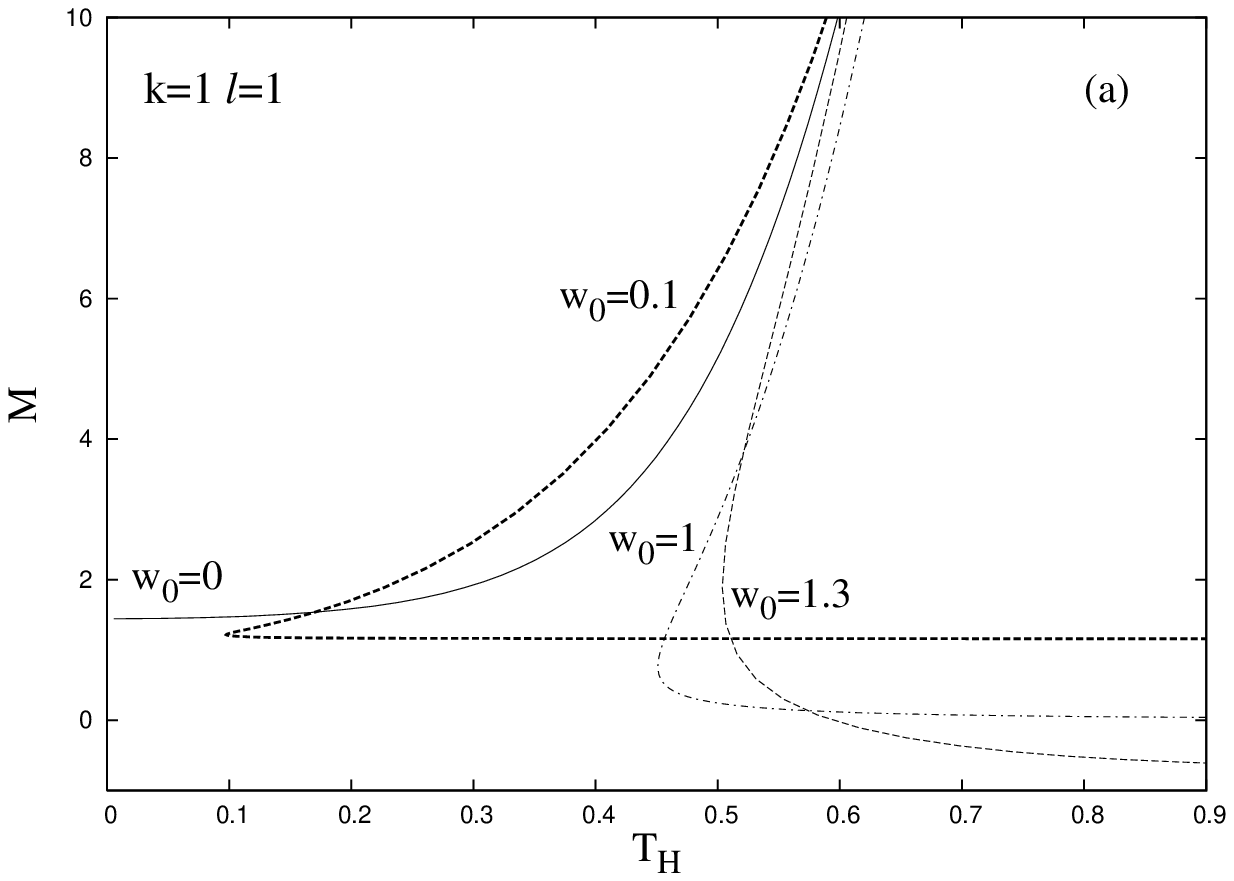}
\includegraphics[width=87mm,angle=0,keepaspectratio]{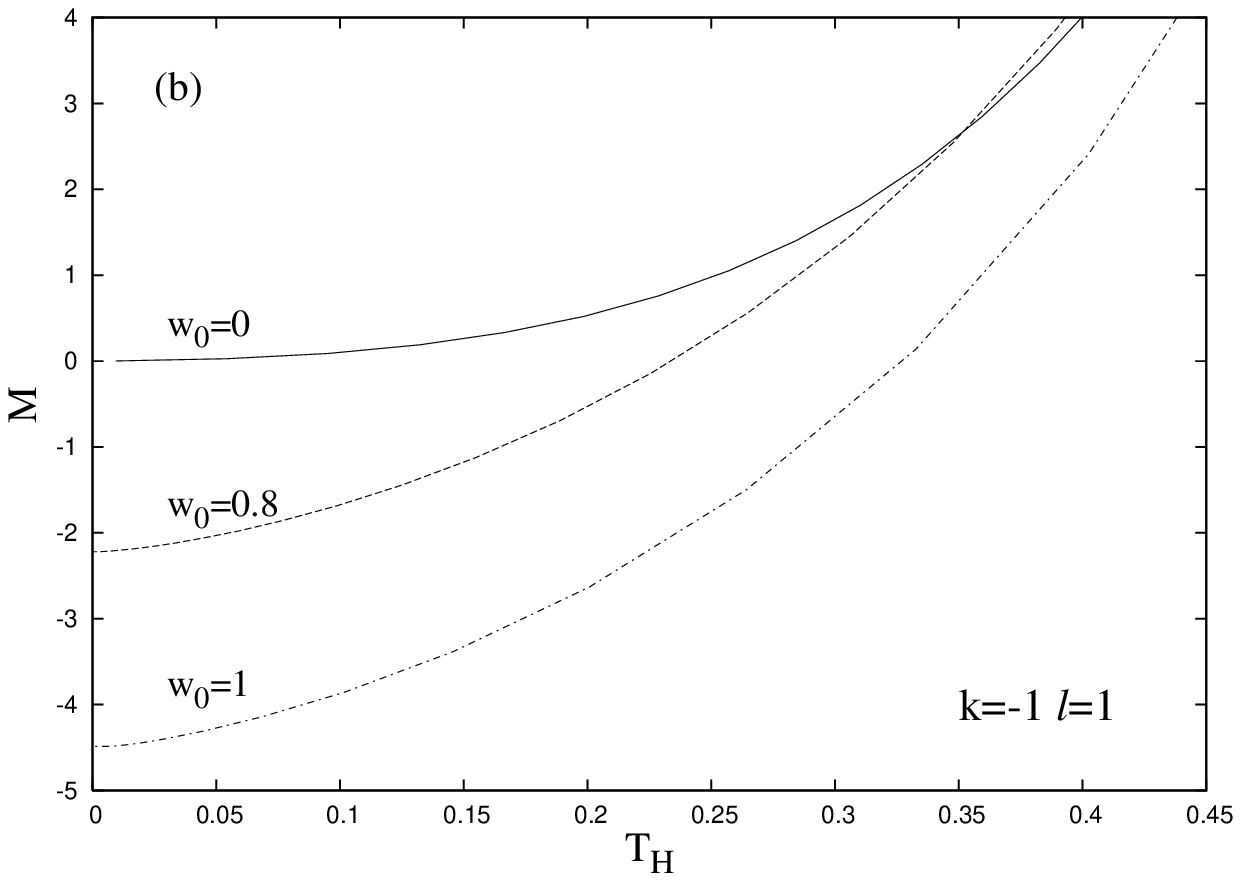}
}}}
\caption{{\small  The mass-parameter $M$ is plotted as a function of temperature
for $k=1,-1$ black hole   solutions and several values
of the magnetic potential at infinity.}}
\end{figure}
%
For $k=\pm 1$, the EYM equations have a nontrivial
 exact solution  \cite{Okuyama:2002mh}
\begin{eqnarray}
\label{ex-sol}
N(r) =k+\frac{r^2}{\ell^2}-\frac {M+8\pi G (k^2/e^2)\log r}{r^2}~,
~~
\sigma(r)=1~,~~\omega(r) = 0,
\end{eqnarray}
which retains the basic features of the general 
configurations.
Solutions  with
a nonvanishing $w(r)$ are constructed 
numerically, the $k=1$ case being considered in \cite{Okuyama:2002mh}
(in the numerics we set $4\pi G/e^2=1$).
%
%
As $r \to \infty$, the spacetime is locally isometric to AdS spacetime,
and we find  the following asymptotic expression of the solutions
(with $M,~w_0,~w_2$ arbitrary parameters\footnote{
By using similar techniques to those employed in the globally regular case
\cite{Okuyama:2002mh}, 
one can prove the absence of non-Abelian black hole solutions with 
$w_0^2=k$.})
\begin{eqnarray}
\nonumber
&N(r) = k+\frac{r^2}{\ell^2}-\frac {M}{r^2} 
-\frac{8\pi G}{e^2}\frac{(w_0^2-k)^2}{r^2}\log(\frac{r}{\ell})+\dots,
~
\sigma(r)  = 1-\frac{16\pi G}{3e^2}\ell^4 w_0^2\frac{(w_0^2-k)^2}{r^6} 
\log^2(\frac{r}{\ell}) +\dots,
\\
\label{asymptBH}
&w (r) = w_{0 } +\frac {w_2}{r^2}-\frac{\ell^2}{r^2}w_0(w_0^2-k)\log (\frac{r}{\ell}) 
+\dots~.
\end{eqnarray}
For all considered values of $(\Lambda,~r_h)$, we find regular black hole solutions
for only one interval $0\leq w_h< w_h^c$. 
The spherically symmetric black holes with $w\neq 0$ have 
a nontrivial globally regular limit $r_h\to 0$.
In contrast, the topological black holes
possess minimal event horizon radius,
for any $w_0$. 
An extremal black hole
is found for the $w(r)=0$ solution (\ref{ex-sol})
with $r_h^2=\ell^2(-k+|k|\sqrt{32 \pi G/(e^2\ell^2)+1})/4$, 
the  parameter $M$ being also fixed by the value of the
cosmological constant.

The action and mass of the AdS$_5$ non-Abelian configurations
is computed by using a boundary
counterterm prescription.
As found in \cite{Balasubramanian:1999re},
the following counterterms are sufficient to cancel
divergences in five dimensions,
for SAdS black hole solution:
\begin{eqnarray}
\label{ct}
I_{\rm ct}=-\frac{1}{8 \pi G} \int_{\partial {\cal M}_r}d^{4}x\sqrt{-h}\Biggl[
\frac{3}{ \ell}+\frac{ \ell}{4}\rm{R}
\Bigg]\ ,
\end{eqnarray}
with $\rm{R}$ the Ricci scalar for 
the boundary metric $h$.
However,  in the presence of matter fields, additional counterterms
may be needed to regulate the action
\cite{Taylor-Robinson:2000xw}.
This is the case  
for the non-Abelian solutions 
discussed in this paper, whose total action (where we have included also 
the Gibbons-Hawking boundary term \cite{Gibbons:1976ue}) diverges logarithmically,
$I=V_k\left(\frac{3 \beta  }{16 \pi G}(M+\frac{k^2\ell^2}{4})-\frac{1}{4G}r_h^3\right)
+\frac{3\beta V_k}{2e^2}(w_0^2-k)^2 \log(\frac{r}{\ell}),
$
(with $V_k$ the area of the surface $\Sigma$).
This divergence is cancelled by a supplementary  counterterm of the form
(with $a,b$ boundary indices):
\begin{eqnarray}
\label{Ict-mat}
I_{ct}^{YM}=
- 
\log(\frac{r}{\ell})
 \int_{\partial {\cal M}_r}d^{4}x\sqrt{-h }\frac{\ell}{2e^2}
 ~{\rm tr}\{ F_{ab}F^{ab}\}~.
\end{eqnarray}
Using these counterterms, one can
construct a divergence-free boundary stress tensor $T_{ab}$ 
\begin{eqnarray}
\label{TAB-mat}
 T_{ab}= 
\frac{1}{8\pi G}(K_{ab}-Kh_{ab}-\frac{3}{\ell}h_{ab}+\frac{\ell}{2} E_{ab})
-\frac{2\ell}{e^2} \log(\frac{r}{\ell})
~{\rm tr}\{F_{ac}F_{bd}h^{cd}-\frac{1}{4}h_{ab}F_{cd}F^{cd}\}~,~~{~} 
\end{eqnarray}
where $E_{ab}$ and $K$ are the Einstein tensor 
 and the trace of the extrinsic curvature $K_{ab}$ for the induced
metric of the boundary, respectively.
In this approach, 
the mass ${\cal M}$ 
of the solutions is the conserved charge associated with the Killing vector 
$\partial /\partial t$  \cite{Balasubramanian:1999re}:
\begin{eqnarray}
\label{mass}
{\cal M}=\frac{3V_k M}{16\pi G}+M_c^{(k)}, ~~{\rm with ~~}M_c^{(k)}=
\frac{3k^2V_k \ell^2}{64\pi G}~.
\end{eqnarray}
We have found that ${\cal M}$
coincides with the  mass computed from the first law of thermodynamics, up to the constant
term $M_c^{(k)}$ which is usually interpreted as the mass of the pure global AdS$_5$.

Based on these results, one can discuss the thermodynamics of 
the non-Abelian black hole solutions in a canonical ensemble,
holding the temperature $T_H$ and the magnetic potential at the boundary at
infinity ($i.e.$ the "magnetic charge") fixed.
Upon application of the Gibbs-Duhem relation $S=\beta {\cal M}-I$, one finds
that the entropy $S$  of these solutions is one quarter of the event horizon area.
The response function whose sign determines the thermodynamic
stability is the heat capacity
$C=  ( {\partial {\cal M}}/{\partial T_H})_{w_0}.$
In Figure 1 we plot the $M(T_H)$ curves for several values of $w_0$ for spherically symmetric
and hyperbolic
 black holes with $\ell=1$
(the results for $k=0$ are rather similar 
to the $k=-1$ case).
For spherically symmetric black holes
with $w_0\neq 0$, the usual SAdS behaviour 
(corresponding to the $w_0=1$ curve in Figure 1a)
is reproduced: the curves
first decrease
toward a minimum, corresponding to the  branch of small unstable
black holes, then increase along the branch of large stable black holes.
The $w(r)=0$ solutions are rather special, since $C>0$ in this case for any $r_h$.
As seen in Figure 1b,
the heat capacity is always positive
for AdS$_5$ non-Abelian topological  black holes.
As a result, the $k=0,-1$ black hole
 solutions
are always thermodynamically locally stable.

From the AdS/CFT correspondence, we expect 
the non-Abelian hairy black holes to be described by
some thermal states
in a dual theory  
 formulated in a metric background given by 
$\gamma_{ab}dx^a dx^b=-dt ^2+\ell^2\left(d\psi^{2}+f^{2}_k(\psi) 
(d \theta^2+\sin^2\theta d \varphi^2)\right).$
One should also consider the interaction of the matter
fields in the dual CFT
with a background
non-Abelian field, whose expression, as read from  (\ref{A}),
(\ref{asymptBH}) is
\begin{equation}
\label{A0}
A_{(0)}=\frac{1}{2} \Big\{ 
\tau_3(\omega_0 d \psi +\cos \theta d \varphi)
-\frac{d f_k(\psi)}{d \psi}(\tau_2 d \theta+\tau_1 \sin
\theta d \varphi) +\omega_0f_k(\psi)(\tau_1 d \theta-\tau_2 \sin
\theta d \varphi)
 \Big\}.
\end{equation}
The expectation value $<\tau^{a}_b>$ of the dual CFT stress tensor
can be calculated using the  relation \cite{Myers:1999qn}
$\sqrt{-\gamma}\gamma^{ab}<\tau_{bc}>=
\lim_{r \rightarrow \infty} \sqrt{-h} h^{ab}{  T}_{bc}.$
Employing also (\ref{TAB-mat}), we find the finite and covariantly 
conserved stress tensor
(with $x^1=\psi,~x^2=\theta,~ x^3=\varphi,~x^4=t$)
\begin{eqnarray}
\label{st1}
8 \pi G <\tau^{a}_b> = 
\frac{1}{2\ell}
\big( 
 \frac{M}{\ell^2} +\frac{k^2}{4}
\big)\left( \begin{array}{cccc}
1&0&0&0
\\
0&1&0&0
\\
0&0&1&0
\\
0&0&0&-3
\end{array}
\right)
-\frac{4\pi G(w_0^2-k)^2 }{e^2\ell^3}
\left( \begin{array}{cccc}
1&0&0&0
\\
0&1&0&0
\\
0&0&1&0
\\
0&0&0&0
\end{array}
\right)
.
\end{eqnarray}
Different $e.g.$ from the case of Reissner-Nordstr\"om-AdS
Abelian solutions, this stress tensor has a nonvanishing trace, 
$ <\tau^{a}_a>={\cal A}_{YM}=-3(w_0^2-k)^2/(2\ell^2 e^2)$.
This agrees with the general results \cite{Coleman:1970je}, \cite{Blau:1999vz},
 \cite{Taylor-Robinson:2000xw} on the trace anomaly in the
presence of an external gauge field, 
${\cal A}_{YM}={\cal R} F_{(0)}^2$,
the coefficient ${\cal R}$ being related to the charges of the fundamental
constituent fields in the dual CFT.

\noindent{\textbf{Non-Abelian black strings solutions.--~}}For 
the situation discussed above, the gravitational
Weyl anomaly ${\cal A}_g$ vanishes, since
 ${\cal A}_g=-\frac{\ell^3}{8\pi G}\left(-\frac{1}{8}
\mathsf{R}_{ab}\mathsf{R}^{ab}+\frac{1}{24}\mathsf{R}^2\right) $
is zero for the induced metric of the boundary.
Here we present an example of configurations where both types of anomalies are present.
This occurs for the non-Abelian version of a class of solutions 
recently considered in \cite{Copsey:2006br,Mann:2006yi} and
describing AdS$_5$ black strings and vortices.
The metric ansatz in this case reads
\begin{eqnarray}
\label{metric-gen-BS} 
ds^{2}= 
 \frac{dr^2}{p(r)}+ r^2 d\Omega^2_{2,k}+a(r)dz^2- b(r)dt^{2},
\end{eqnarray}
where $d\Omega^2_{2,k}=d\theta^{2}+f^{2}_k(\theta) d\varphi^2$ denotes the line
element of a two-dimensional space with
constant curvature, and the direction $z$ is periodic with period $L$.
Considering again an SU(2) YM field, 
the gauge field ansatz has two magnetic potentials and reads
\begin{eqnarray}
\label{YM-BS} 
A=\frac{1}{2} \bigg \{  
  \omega(r) \tau_1    d \theta
+\left(\frac{d \ln f_k(\theta)}{d \theta} \tau_3
+ \omega(r) \tau_2  \right) f_k(\theta) d \varphi +H(r)\tau_3 dz \bigg \}~. 
\end{eqnarray} 
Similar to the black hole case, 
we have found a continuum of black string solutions presenting an event horizon
at $r=r_h$, where
$p(r_h)=b(r_h)=0$, while $a(r_h)=a_h>0$, $w(r_h)=w_h$, $H(r_h)=H_h$.
The Hawking temperature of the black strings is $T_H= {\sqrt{b'(r_h)p'(r_h)}}/{4\pi}.$
%
\begin{figure}[h!]
\parbox{\textwidth}
{\centerline{
\mbox{
\epsfysize=10.0cm
\includegraphics[width=92mm,angle=0,keepaspectratio]{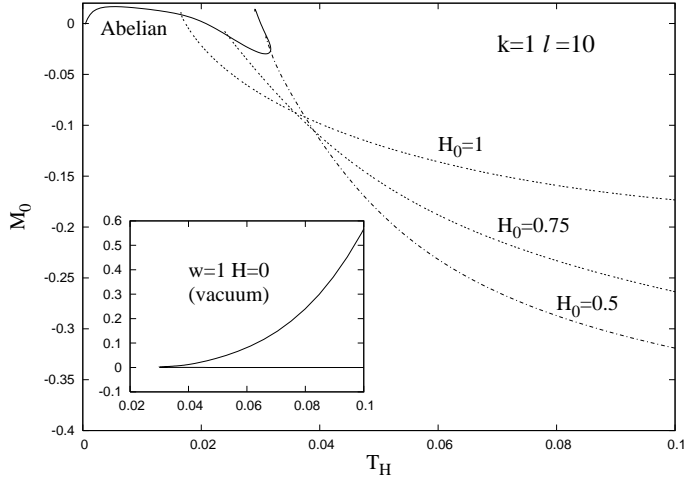} 
}}}
\caption{{\small
The mass-parameter $M_0$ is plotted for $k=1$ black string solutions.}}
\end{figure} 
%
%
The solutions have the following asymptotic expression in 
terms of four arbitrary constants $c_t,~c_z,~H_0$ and $w_2$:
\begin{eqnarray}
\nonumber
&&a(r)=\frac{k}{2}+\frac{r^2}{\ell^2}+c_z(\frac{\ell}{r })^2+
\frac{k^2}{2}(\frac{1}{6}-\frac{8\pi G}{e^2\ell^2})
\log \frac{r}{\ell} (\frac{\ell}{r })^2+\dots,
\\
\label{asympt-BS} 
&&b(r)=\frac{k}{2}+\frac{r^2}{\ell^2}+c_t(\frac{\ell}{r })^2+
\frac{k^2}{2}(\frac{1}{6}-\frac{8\pi G}{e^2\ell^2})
\log \frac{r}{\ell} (\frac{\ell}{r })^2+\dots,
\\
\nonumber
&&p(r)=\frac{2k}{3}+\frac{r^2}{\ell^2}+(c_t+c_z+\frac{8\pi G}{e^2\ell^2})(\frac{\ell}{r })^2+
k^2(\frac{1}{6}-\frac{8\pi G}{e^2\ell^2})
\log \frac{r}{\ell} (\frac{\ell}{r })^2+\dots,
\\
\nonumber
&&w(r)=\frac{w_2}{r^2}+\dots, ~~H(r)=H_0(1+\frac{w_2^2\ell^2}{12 r^6})+\dots~.
\end{eqnarray}
The basic features of the black strings are  similar to the black hole case.
Again, the $k=1$ solutions possess nontrivial
globally regular limits, representing the AdS counterparts
of the $\Lambda=0$ non-Abelian vortices in Ref. \cite{Volkov:2001tb}.
The $k=0,-1$ topological black strings present a minimal event horizon
radius.
For given $(r_h,\Lambda)$ the solutions' global charges depend
on the value of the magnetic
gauge potential $H$ at infinity, which is a free parameter.
The solutions with  $w(r)=0$, $H(r)=const.$ represent Abelian black strings,
generalizing the exact BPS solutions in \cite{Klemm:2000nj}.
These configurations exist for values of the event horizon radius
greater than a minimal value $r_{h}^c$, an
extremal solution being approached in that limit.
The non-Abelian solutions  depend
on the value $H_0$ and exist on a finite interval of $r_h$.
In the limit $r_h \to r_{h}^c$ the gauge function $w(r)$ vanishes identically and
the branch of non-Abelian solutions bifurcates into the Abelian
 branch.

The action and global charges of these configurations
are computed by employing again  the counterterm formalism.
As found in \cite{Mann:2006yi} 
the action of the vacuum solutions presents a logarithmic divergence
which is regularized by adding the following term to the boundary action
\cite{Skenderis:2000in}:
\begin{eqnarray}
I_{\mathrm{ct}}^{s} &=&\frac{1}{8\pi G }\log(\frac{r}{\ell}) 
\int_{\partial {\cal M}_r} d^{4}x\sqrt{-h }
 \frac{\ell^3}{8}(\frac{1}{3}\mathsf{R}^2-\mathsf{R}_{ab}\mathsf{R}^{ab})~,
\end{eqnarray}
which implies a supplementary contribution to the boundary stress tensor
(\ref{TAB-mat}).
The bulk YM fields give another logarithmic divergence, 
which is regularized by the matter counterterm (\ref{Ict-mat}). 
As usual with black strings \cite{Harmark:2007md},
apart from mass ${\cal M}$, there is also 
a second global charge associated with the Killing vector
$\partial/\partial z$ and corresponding 
to the solutions' tension ${\mathcal T}$:
\begin{eqnarray}
\label{MT} 
{\cal M}&=&M_0+M_c^{(k )}~,~~M_0=\frac{\ell LV_{k}}{16\pi G}\big[c_z-3 c_t\big]~,
\\
\nonumber
{\mathcal T}&=&{\mathcal T}_0+{\mathcal T}_c^{(k)}~,~~
{\mathcal T}_0=\frac{\ell V_{k }}{16\pi G }\big[3c_z-c_t\big] 
~,
{\rm with~~}M_c^{(k )}=L{\mathcal T}_c^{(k)} =\frac{\ell }{16\pi G}V_{k }L,
\end{eqnarray} 
%
where $V_{k }$ is the total area of the angular sector,
$M_c^{(k )}$ and ${\mathcal T}_c^{(k )}$ being  Casimir-like terms.
In Figure 2 we plot the mass-parameter
$M_0$ as a function of temperature for $k=1$ black strings
with several values of $H_0$ (in a $d=4$ picture, this corresponds to
different  vacuum expectation values of the Higgs field \cite{Volkov:2001tb}).
One can see that, in contrast with the vacuum case, 
the non-Abelian black strings are thermally unstable.
The situation is more complicated in the Abelian case, 
the solutions near extremality possessing a positive 
heat capacity.
 
For these black strings solutions,
the background metric upon which the dual field theory resides is
$\gamma_{ab}dx^adx^b=-dt^2+dz^2+\ell^2(d\theta^{2}+f^{2}_k(\theta) d\varphi^2)~.$
The boundary CFT is formulated in this case
in a   background Abelian gauge field, with
\begin{eqnarray}
\label{YM-BS-bound} 
A_{(0)}=\frac{\tau_3}{2} \bigg \{  
   \frac{d  f_k(\theta)}{d \theta}  d \varphi +H_0  dz \bigg \}~. 
\end{eqnarray} 
The expectation value of the stress tensor of the dual CFT contains four different parts
(with $x^1=\theta,~x^2=\varphi,~ x^3=z,~x^4=t$)
\begin{eqnarray}
\label{st2}
8 \pi G <\tau^{a}_b>~= 
 -
\frac{c_z}{2 \ell  } \left( \begin{array}{cccc}
1&0&0&0
\\
0&1&0&0
\\
0&0&-3&0
\\
0&0&0&1
\end{array}
\right)
-
\frac{c_t}{ 2 \ell } 
\left( \begin{array}{cccc}
1&0&0&0
\\
0&1&0&0
\\
0&0&1&0
\\
0&0&0&-3
\end{array}
\right)
\\
\nonumber
+\frac{k^2}{24 \ell }
\left( \begin{array}{cccc}
2&0&0&0
\\
0&2&0&0
\\
0&0&-1&0
\\
0&0&0&-1
\end{array} 
\right)
-\frac{2\pi G}{e^2 \ell^3} k^2
\left( \begin{array}{cccc}
1&0&0&0
\\
0&1&0&0
\\
0&0&0&0
\\
0&0&0&0
\end{array}
\right)
.
\end{eqnarray}
 The  trace of this tensor is 
equal to the sum of the gravitational and external gauge field contributions
${\cal A}={\cal A}_g+{\cal A}_{YM}=k^2(\frac{1}{96\pi G \ell}-\frac{1}{2e^2\ell^3})$,
vanishing for the Abelian BPS solutions in \cite{Klemm:2000nj}.

\noindent{\textbf{Further remarks.--~}}On general grounds, one expects that extending 
the known classes of solutions of the $d=5$ supergravity
to a non-Abelian gauge group would lead to a variety of new physical
effects. 
The black objects discussed in this paper are perhaps the simplest solutions
relevant in this context.
We expect a much richer structure to be found when relaxing the spacetime symmetries,
or when taking a more general gauge group.
However, the generic non-Abelian solutions will always present a nonvanishing
magnetic gauge field  on the boundary which appears
as a background for the dual theory.
Also, similar to the $d=4$ case \cite{Mann:2006jc}, 
the existence of both spherically symmetric globally regular
and hairy black hole solutions
with the same set of data at infinity raises the question as to how 
the dual CFT is able to distinguish between these different bulk configurations. 
\\
\\
{\bf\large Acknowledgements} \\
YB is grateful to the
Belgian FNRS for financial support.
The work of ER is carried out
in the framework of Enterprise--Ireland Basic Science Research Project
SC/2003/390 of Enterprise-Ireland.

\begin{small}

\end{small}

\end{document}